\documentclass[%
 reprint,twocolumn,floatfix,
 amsmath,amssymb,
 aps,
 prl,
superscriptaddress,
ctex
]{revtex4}

\usepackage{natbib}
\usepackage{graphicx}% Include figure files
\usepackage{dcolumn}% Align table columns on decimal point
\usepackage{bm}% bold math
\usepackage{graphicx,color}
\usepackage{comment}
\usepackage{url}
\usepackage[normalem]{ulem}
\usepackage{CJK}
\usepackage[colorlinks=true, linkcolor=red, citecolor=blue]{hyperref}

\begin{document}

\title{Dark energy and matter interacting scenario to relieve $H_0$ and $S_8$ tensions}
% Force line breaks with \\

\author{Li-Yang Gao}\thanks{These authors contributed equally to this paper.}
\affiliation{Key Laboratory of Cosmology and Astrophysics (Liaoning Province) \& Department of Physics, College of Sciences, Northeastern University, Shenyang 110819, China}%
\affiliation{Kapteyn Astronomical Institute, University of Groningen, P.O. Box 800, 9700 AV Groningen, the Netherlands}%
\author{She-Sheng Xue}\thanks{These authors contributed equally to this paper.}
\affiliation{ICRANet, Piazzale della Repubblica, 10-65122, Pescara, Italy}%
\affiliation{ ICTP-AP, University of Chinese Academy of Sciences, Beijing, China}%
\affiliation{Physics Department, Sapienza University of Rome, P.le A. Moro 5, 00185, Rome, Italy}%
\affiliation{INFN, Sezione di Perugia, Via A. Pascoli, I-06123, Perugia, Italy}%
\author{Xin Zhang}
\thanks{Corresponding author. \\zhangxin@mail.neu.edu.cn}%
\affiliation{Key Laboratory of Cosmology and Astrophysics (Liaoning Province) \& Department of Physics, College of Sciences, Northeastern University, Shenyang 110819, China}%
\affiliation{Key Laboratory of Data Analytics and Optimization for Smart Industry (Ministry of Education), Northeastern University, Shenyang 110819, China}%
\affiliation{National Frontiers Science Center for Industrial Intelligence and Systems Optimization, Northeastern University, Shenyang 110819, China}%

\begin{abstract}
We consider a new cosmological model (called $\tilde\Lambda$CDM) in which the vacuum energy interacts with matter and radiation, and test this model using the current cosmological observations. Using the CMB+BAO+SN (CBS) dataset to constrain the model, we find that $H_0$ and $S_8$ tensions are relieved to $2.87\sigma$ and $2.77\sigma$, respectively. However, in this case, the $\tilde\Lambda$CDM model is not favored by the data, compared with $\Lambda$CDM. We find that when the $H_0$ and $S_8$ data are added to the data combination, the situation is significantly improved. In the CBS+$H_0$ case, the model relieves the $H_0$ tension to $0.47\sigma$, and the model is favored over $\Lambda$CDM. In the CBS+$H_0$+$S_8$ case, we obtain a synthetically best situation, in which the $H_0$ and $S_8$ tensions are relieved to $0.72\sigma$ and $2.11\sigma$, respectively. In this case, the model is most favored by the data. Therefore, this cosmological model can greatly relieve the $H_0$ tension and simultaneously effectively alleviate the $S_8$ tension.

\end{abstract}

\maketitle

%%%%%%%%%%%%%%%%%%%%%%%%%%%%%%%%%%%%%%%%%%%%%%%%%%%%%%%%%%%%%%%%%%%%%%%%

%\noindent{\bf \em Introduction.}  
{\it Introduction.---}The discovery of the accelerating expansion of the universe in 1998 \cite{SupernovaSearchTeam:1998fmf, SupernovaCosmologyProject:1998vns} significantly invigorated interest in cosmology. Subsequent observations have continually supplied increasingly precise data, enhancing our understanding of this phenomenon. Key contributions include the cosmic microwave background (CMB) data from {\it Planck} 2018 \cite{Planck:2018vyg} and the direct measurement of the Hubble constant, $H_0$ \cite{Riess:2021jrx}. The enhanced precision of these observational data has led to the emergence of $H_0$ and $S_8$ tensions as prominent new challenges in the field.

%Greater zeal for cosmology resulted from the 1998 finding of the universe's accelerated expansion \cite{SupernovaSearchTeam:1998fmf, SupernovaCosmologyProject:1998vns}. More and more new observations have been providing ever-more-accurate data that can be used to comprehensively understand the cause of the universe's accelerated expansion, such as the cosmic microwave background (CMB) from {\it Planck} 2018 \cite{Planck:2018vyg} and the direct measurement of the Hubble constant $H_0$ \cite{Riess:2021jrx}. Due to the increasing accuracy of observational data, $H_0$ and $S_8$ tensions have emerged as new issues.

The early-universe observation of {\it Planck} TT, TE, and EE + lowE + lensing \cite{Planck:2018vyg} combined with baryon acoustic oscillation (BAO) measurements from galaxy redshift surveys \cite{Beutler:2011hx, Ross:2014qpa, BOSS:2016wmc} gives a fit result of $H_0=67.36 \pm 0.54~\rm{km~s^{-1} Mpc^{-1}}$ for the $\Lambda$CDM model. The late-universe observation of the Cepheid-type Ia supernova (SN) distance ladder by SH0ES \cite{Riess:2021jrx} gives a result of $H_0=73.04 \pm 1.04~\rm{km~s^{-1} Mpc^{-1}}$. Thus, the $H_0$ tension has currently reached $4.85\sigma$ \cite{DiValentino:2021izs,Verde:2019ivm,DiValentino:2017gzb,DiValentino:2020zio,Freedman:2017yms,Riess:2019qba}.
In addition, the result of $S_8=0.832 \pm 0.013$ obtained from {\it Planck} $2018$ is in a $3.08\sigma$ tension with the result of $S_8= 0.766^{+0.020}_{-0.014}$ obtained from the combination of KiDS/Viking and SDSS cosmic shear data \cite{Heymans:2020gsg}.
As the precision of the data increases, the $H_0$ and $S_8$ tensions become more pronounced.
Consequently, the issue is unlikely caused by the accuracy of the data but by issues with the measurements or $\Lambda$CDM model.

Numerous studies have attempted to examine the systematic flaws in both methodologies to address these tensions (see, for example, Refs.~\cite{Spergel:2013rxa, Addison:2015wyg, Planck:2016tof, Efstathiou:2013via, Cardona:2016ems, Zhang:2017aqn, Follin:2017ljs}); however, no conclusive evidence has been found.
As a result, new independent measurement techniques for late-universe observation have drawn attention, including, for example, the substitution of Mira variables \cite{Huang:2019yhh} or red giants \cite{Yuan:2019npk} for Cepheids in the Cepheid-SN distance ladder, observation of strong lensing time delays \cite{Wong:2019kwg}, water masers \cite{Pesce:2020xfe}, surface brightness fluctuations \cite{Jensen:1998bi}, gravitational waves from neutron star mergers \cite{LIGOScientific:2017adf}, use of the different ages of galaxies as cosmic clocks \cite{Jimenez:2001gg, Moresco:2016mzx}, and the baryonic Tully-Fisher relation \cite{Schombert:2020pxm}.
Despite numerous new observations, the issue of the $H_0$ tension has not yet been resolved.

In this study, we assume that both early- and late-universe observations are credible and that the cosmological model must be modified.
The $\Lambda$CDM model has undergone numerous adaptations, which can be divided into two main groups: the dark energy and modified gravity (MG) models.
Numerous dark energy models have been used to reduce the $H_0$ tension \cite{Linder:2002et, Grande:2006nn, Dutta:2018vmq, Akarsu:2019hmw, Ye:2020btb, Vazquez2012, Calderon:2020hoc,Guo:2018ans, Guo:2021rrz, Gao:2021xnk, Zhao:2017urm, Zhang:2015uhk, Zhang:2014dxk, Feng:2019jqa, Li:2013dha, Hill:2020osr, Ivanov:2020ril, Smith:2020rxx, Smith:2019ihp, Agrawal:2019lmo, Alexander:2019rsc, Niedermann:2019olb, Kaloper:2019lpl}, and some researchers have explored modifications in gravitational theories, referred to as MG models \cite{Boisseau2015, Braglia:2020iik, Ballesteros:2020sik, Ballardini:2020iws, Zumalacarregui:2020cjh}.
Although these models partially mitigate the $H_0$ tension, they often exacerbate the $S_8$ tension \cite{Guo:2018ans}.

{Using the concepts of asymptotic safety \cite{Weinberg2009} and particle production \cite{PhysRevD.7.2357} in gravitational field theory, Ref. \cite{Xue:2014kna} introduced a $\tilde\Lambda$CDM model characterized by a dynamical vacuum energy component $\tilde\Lambda$ that interacts with matter and radiation.
The density of vacuum energy, $\rho_{\Lambda}(t)=\tilde\Lambda/(8\pi G)$, undergoes conversion to matter (radiation) and {\it vice versa}.
The $\tilde\Lambda$CDM model has been employed to investigate various cosmological phenomena, including inflation \cite{Xue:2021jyj}, reheating \cite{Xue:2020tpf}, and the evolution of standard cosmology from the end of reheating to the current era \cite{Xue:2023qft, Xue2022}.
}
% \textcolor{blue}{Inspired by the asymptotic safety \cite{Weinberg2009} and particle production \cite{PhysRevD.7.2357} of gravitational field theory, Ref.~\cite{Xue:2014kna} presented a $\tilde\Lambda$CDM model of a time-varying dark energy $\tilde\Lambda$ interacting with matter and radiation. 
% The dark-energy density $\rho_{_\Lambda}(t)=\tilde\Lambda/(8\pi G)$ converts to matter (radiation) and {\it vice versa}. The $\tilde\Lambda$CDM model is applied to study inflation \cite{Xue:2021jyj}, reheating \cite{Xue:2020tpf} and standard cosmology from the reheating end to the present time \cite{Xue:2023qft, Xue2022}.}
Reference~\cite{Begue2019} presents a preliminary phenomenological study of the $\tilde\Lambda$CDM model, comparing it with $\Lambda$CDM and other interacting dark energy models. Reference~\cite{Gao:2021xnk} presents the investigation of a simplified $\tilde\Lambda$CDM model, showing the possibility of greatly relieving the $H_0$ tension. 
Here, we further study the $H_0$ and 
$S_8$ tensions and their correlations within the entire parameter space of the $\tilde\Lambda$CDM model.

To verify whether the model alleviates both the $H_0$ and $S_8$ tensions, we adopt the most commonly used data combination, $\rm CMB+BAO+SN$ (CBS).
We also consider the $H_0$ and $S_8$ measurements as complementary to CBS.
Note that the redshifts of the galaxies used for measuring $S_8$ are generally less than $1.5$.
We find that the model relieves the $H_0$ tension to less than $1\sigma$ and simultaneously relieves the $S_8$ tension, by simply using the CBS data. 
The lowest $H_0$ and $S_8$ tensions are achieved using the CBS$+H_0$ and CBS$+H_0+S_8$ data, respectively.
Although the $S_8$ tension is still slightly greater than $2\sigma$, we show that it is definitely reduced when the $H_0$ tension is reduced. This situation is difficult to observe in other cosmological models.

%%%%%%%%%%%%%%%%%%%%%%%%%%%%%%%%%%%%%%%%%%%%%%%%%%%%%%%%%%%%%%%%%%%%%%%%

%\noindent{\bf \em Motivation \& Theoretical Model.} 
{\it Motivation and theoretical model.---}
In the framework of the $\tilde\Lambda$CDM model in standard cosmology, particles are created in the neighborhood of the Friedmann universe horizon. Such dynamics is effectively described by the particle production rate $\Gamma_m=-(\chi/4\pi)(\dot H/H^2)$ and energy density $\rho_{m}^H=2 \chi m^2H^2$ of produced particles. Here, $m$ represents the particle mass, and $\chi$ characterizes the width $\chi/m$ of the particle-production layer on the horizon. The interplay between vacuum energy and matter (radiation) is delineated through the particle production rate $\Gamma_m$ and density $\rho_{m}^H$, explicated in Refs. \cite{Xue:2020tpf, Xue:2023qft} as
\begin{eqnarray}
&&H^2 = \frac{8\pi G}{3}(\rho_{m}+\rho_{r}+\rho_{\Lambda}),\label{friedman0}\\
&&\dot H =-\frac{4\pi G}{3}(3\rho_{m}+4\rho_{r}),
\label{friedman}\\
&&\dot\rho_{m}+ 3H\rho_{m} = \Gamma_m(\rho_{m}^H - \rho_{m}-\rho_{{r}}),
\label{rateeqd}\\
&&\dot\rho_{{r}}+ 4 H\rho_{{r}} = \Gamma_m(\rho_{m}^H - \rho_{{m}}-\rho_{{r}}).
\label{rateeqr}
\end{eqnarray}

The matter density $\rho_{{m}}$
is for non-relativistic particles, and the radiation density $\rho_{{r}}$ for relativistic particles. 
Equations (\ref{friedman0}) and (\ref{friedman}) are Friedmann equations for the time-varying cosmological term $\rho_{\Lambda}(t)$, and we have $p_{\Lambda}=-\rho_{\Lambda}$. 
Vacuum energy and matter (radiation) interact via the right-handed sides of the energy conservation equations (\ref{rateeqd}) and (\ref{rateeqr}). With initial values at either the reheating end or today, four dynamic equations form a closed set, providing the unique solutions $\rho_{{m,r,\Lambda}}$ and $H$. 

The numerical solutions are too complex to proceed
with data analysis.
Nevertheless, the model accommodates scaling solutions, consistent with the principles of asymptotic safety in gravitational theories \cite{Weinberg2009, Xue:2014kna}.
That is, the $\tilde\Lambda$CDM quantities $\rho_{{m,r,\Lambda}}$ receive scaling factor $(1+z)^\delta$ (with $|\delta|\ll 1$) corrections to 
$\Lambda {\rm CDM}$ counterparts when the redshift $z$ becomes small (late times). Thus, we produce ansatz solutions $\rho_{{m,r}}\propto (1+z)^{3(1+w_{m,r})-\delta^{M,R}_G}$ deviating from normal matter with $w_m=0$ and normal radiation with $w_r= 1/3$, and $\rho_{\Lambda}\propto (1+z)^{3(1+w_{\Lambda})+\delta_{\Lambda}}$ deviating from normal vacuum energy with $w_{\Lambda}=-1$.

Therefore, in late times, the $\tilde\Lambda$CDM model parameterizes the Friedmann equation as 
\begin{equation}
E^2(z)=\Omega_{\rm m}(1+z)^{(3-\delta^M_{\rm G})}+\Omega_{\rm r}(1+z)^{(4-\delta^R_{\rm G})}+\Omega_{_\Lambda}(1+z)^{\delta_\Lambda},
\label{final+}
\end{equation}
where $E(z)=H(z)/H_0$, and the three scaling indexes $\delta^{M,R}_{\rm G}$ and $\delta_{\Lambda}$ are considerably than unity. 
The generalized conservation law yields
\begin{equation}
(1+z)\frac{d}{dz}E^2(z)=3\Omega_{\rm m}(1+z)^{(3-\delta^M_{\rm G})}+4\Omega_{\rm r}(1+z)^{(4-\delta^R_{\rm G})}.
\label{fcgeqi20+}
\end{equation}
At the leading 
order of $\delta^{M,R}_{\rm G}$ 
and $\delta_{\Lambda}$ and for low 
redshifts, we find the relation
\begin{eqnarray}
\delta_\Lambda &\approx & (\Omega_{\rm m}\delta^M_{\rm G}+\Omega_{\rm r}\delta^R_{\rm G})/\Omega_{_\Lambda}.\
 \label{deltal+}
\end{eqnarray}
Two independent parameters $\delta^{M, R}_{\rm G}$ can be constrained by observational data. 
Their negative (positive) values indicate the process of radiation and matter conversion into dark energy (the inverse process).     
% In the simplified $\tilde\Lambda$CDM model
% \cite{Gao:2021xnk}, we assume that $\delta_G\equiv \delta_{_G}^M=\delta_{_G}^R$ and $\delta_\Lambda=\delta_G(\Omega_{\rm m}+\Omega_{\rm r})/\Omega_\Lambda$.
{For a comparative analysis, we also introduce the simplified $\tilde\Lambda$CDM model from our previous study \cite{Gao:2021xnk}, characterized by $\delta_G\equiv \delta_{_G}^M=\delta_{_G}^R$ and $\delta_\Lambda=\delta_G(\Omega_{\rm m}+\Omega_{\rm r})/\Omega_\Lambda$.}
%%%%%%%%%%%%%%%%%%%%%%%%%%%%%%%%%%%%%%%%%%%%%%%%%%%%%%%%%%%%%%%%%%%%%%%%

%\noindent{\bf \em Data \& Methodology.} 
{\it Data and methodology.---}In this study, we employ CMB, BAO, SN, $H_0$, and $S_8$ data.
For CMB data, we use the {\it Planck} 2018 full-mission TT, TE, and EE + lowE + lensing power spectrum \cite{Planck:2018vyg}.
For BAO data, we utilize five points from three observations (the 6dF Galaxy Survey, SDSS DR7 Main Galaxy Sample, and DR12 galaxy sample) \cite{Beutler:2011hx, Ross:2014qpa, BOSS:2016wmc}.
For SN data, we use 1048 data points from the Pantheon compilation \cite{Pan-STARRS1:2017jku}.
In addition, we also consider two Gaussian priors, that is, $S_8= 0.766^{+0.020}_{-0.014}$ [here, $S_8\equiv\sigma_8(\Omega_{\rm m}/0.3)^{0.5}$], which originated from the combination of the KiDS/Viking and SDSS data \cite{Heymans:2020gsg}, and $H_0=73.04 \pm 1.04~\rm{km~s^{-1} Mpc^{-1}}$, which was determined from the distance ladder by the SH0ES team \cite{Riess:2021jrx}.

To conduct the Markov-chain Monte Carlo (MCMC) analysis, we use the {\tt MontePython} code \cite{Audren2012ConservativeCO}. 
To assess how well the various models fit the data, we use the Akaike information criterion (AIC) \cite{Szydlowski:2008by, Huterer:2016uyq,delCampo:2011jp}, ${\rm AIC}\equiv \chi^2 +2d$, where $d$ is the number of free parameters. 
We use $\Delta {\rm AIC}=\Delta\chi^2+2\Delta d$ to compare a model with $\Lambda$CDM.

%%%%%%%%%%%%%%%%%%%%%%%%%%%%%%%%%%%%%%%%%%%%%%%%%%%%%%%%%%%%%%%%%%%%%%%%

\begin{figure}[t]
\includegraphics[trim=0.1cm 0.1cm 0cm 0cm,clip, width=0.45\textwidth]{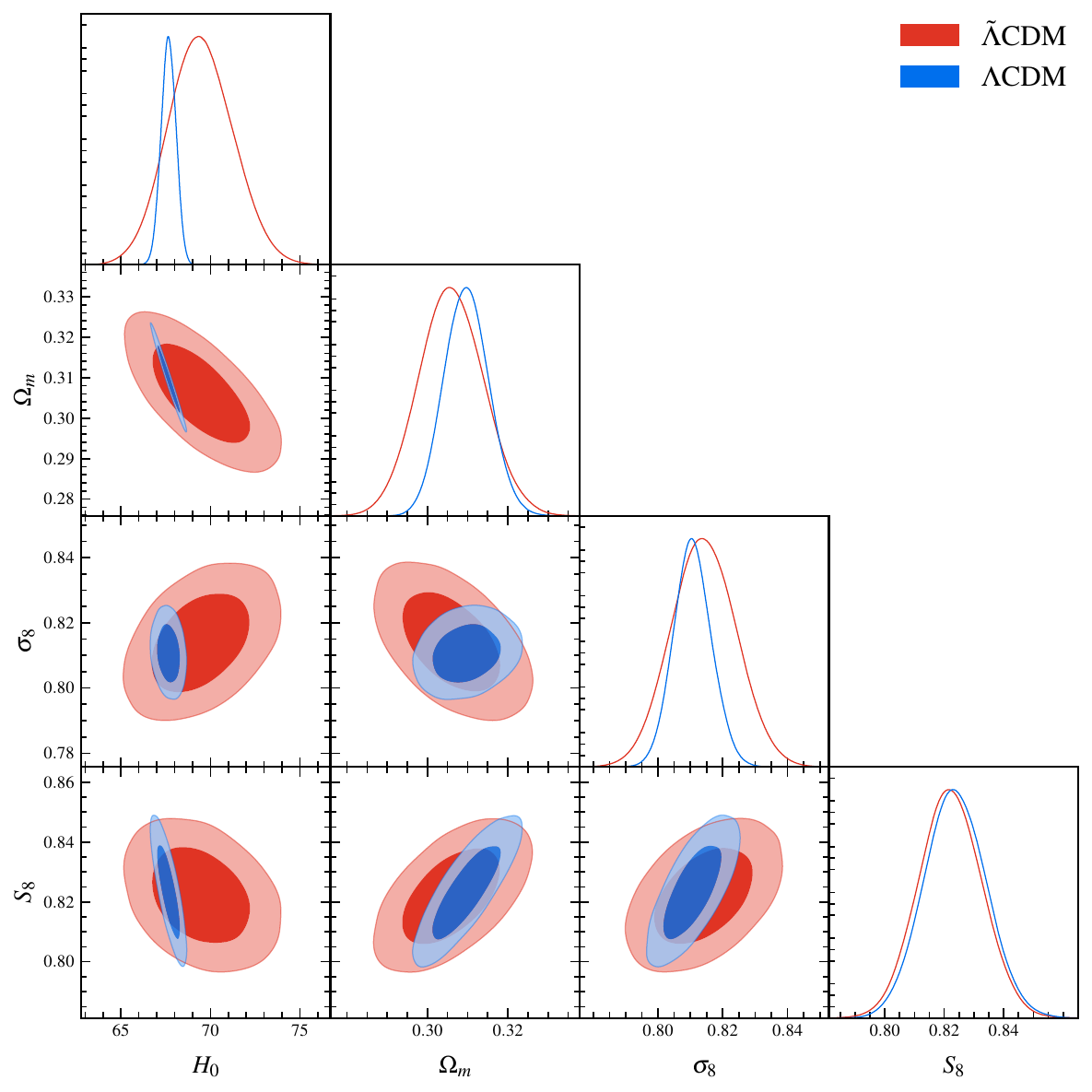}
\caption{Constraints ($68.3\%$ and $95.4\%$ confidence level) on $H_0$, $\Omega_{\rm m}$, $\sigma_8$, and $S_8$ in the $\Lambda$CDM and $\tilde\Lambda$CDM models using the CBS data. Here, $H_0$ is in units of ${\rm km~s^{-1}~Mpc^{-1}}$.}
\label{CBSresults}
\end{figure}

% \begin{table}
% \renewcommand{\arraystretch}{1.3}
% %\centering
% %\caption{The constraint results of parameters in the $\Lambda$CDM, CPL and $\tilde\Lambda$CDM models with the CBS data set.}\label{tab1}
% \begin{tabular}{|c|c c|}%\tiny
% \hline
% Model &$\Lambda$CDM&$\tilde\Lambda$CDM\\
% \hline
% $\delta^M_{\rm G}$&-&$-0.00071^{+0.0011}_{-0.00082}$\\
% $\delta^R_{\rm G}$&-&$-0.0101^{+0.0047}_{-0.0042}$\\
% $\delta_{\Lambda}$&-&$-0.00030^{+0.00045}_{-0.00037}$\\
% $\Omega_{\rm m}$&$0.3097\pm 0.0055$&$0.3038\pm 0.0083$\\
% $H_0~[{\rm km~s^{-1}~Mpc^{-1}}]$&$ 67.66\pm 0.41 $&$70.6^{+1.4}_{-1.7}$\\
% $\sigma_8$&$0.8107\pm 0.0059$&$0.815^{+0.010}_{-0.012}$\\
% $S_8$&$0.824\pm 0.010 $&$0.820\pm 0.011$\\
% \hline
% ${H_0}~{\rm tension}$&$4.81\sigma$&$1.28\sigma$\\
% % ${\sigma_8}~{\rm tension}$&$2.14\sigma$&$2.45\sigma$&$2.31\sigma$\\
% ${S_8}~{\rm tension}$&$2.94\sigma$&$2.67\sigma$\\
% $\chi_{\rm min}^{2}$&$1907.55$&$1909.17$\\
% $\Delta {\rm AIC}$&$0$&$5.62$\\
% \hline
% \end{tabular}
% \centering
% \caption{\red{old} Constraint results of parameters in the $\Lambda$CDM and $\tilde\Lambda$CDM models with the CBS data. Here $H_0$ is in units of ${\rm km~s^{-1}~Mpc^{-1}}$.}\label{tab1}
% \end{table}

\begin{table}
\renewcommand{\arraystretch}{1.3}
%\centering
%\caption{The constraint results of parameters in the $\Lambda$CDM, CPL and $\tilde\Lambda$CDM models with the CBS data set.}\label{tab1}
\begin{tabular}{|c|c c|}%\tiny
\hline
Model &$\Lambda$CDM&$\tilde\Lambda$CDM\\
\hline
$\delta^M_{\rm G}$&-&$-0.00052\pm 0.00088$\\
$\delta^R_{\rm G}$&-&$-0.0061\pm 0.0059$\\
$\delta_{\Lambda}$&-&$-0.00022\pm 0.00038$\\
$\Omega_{\rm m}$&$0.3097\pm 0.0055$&$0.3060\pm 0.0081$\\
$H_0~[{\rm km~s^{-1}~Mpc^{-1}}]$&$ 67.66\pm 0.41 $&$69.5\pm 1.8$\\
$\sigma_8$&$0.8107\pm 0.0059$&$0.814\pm 0.010$\\
$S_8$&$0.824\pm 0.010 $&$0.822\pm 0.011$\\
\hline
${H_0}~{\rm tension}$&$4.81\sigma$&$2.87\sigma$\\
% ${\sigma_8}~{\rm tension}$&$2.14\sigma$&$2.45\sigma$&$2.31\sigma$\\
${S_8}~{\rm tension}$&$2.94\sigma$&$2.77\sigma$\\
$\chi_{\rm min}^{2}$&$1907.55$&$1907.34$\\
$\Delta {\rm AIC}$&$0$&$3.79$\\
\hline
\end{tabular}
\centering
\caption{Constraint results of parameters in the $\Lambda$CDM and $\tilde\Lambda$CDM models with the CBS data. Here $H_0$ is in units of ${\rm km~s^{-1}~Mpc^{-1}}$.}\label{tab1}
\end{table}

%\noindent{\bf \em Results.} 
{\it Results.---}To evaluate the $\tilde\Lambda$CDM model, we use the $\Lambda$CDM model as a reference. 
First, we constrain the $\Lambda$CDM and $\tilde\Lambda$CDM models using the joint CBS data. 
The results are shown in  Table~\ref{tab1} and Fig.~\ref{CBSresults}.
In the $\tilde\Lambda$CDM model, we obtain fit values of {$H_0=69.5\pm 1.8~\rm{km~s^{-1} Mpc^{-1}}$ and $S_8=0.822\pm 0.011$}; thus, the $H_0$ and $S_8$ tensions are relieved to be at the {$2.87\sigma$ and $2.77\sigma$} levels, respectively. 
We find that, in this case, the $H_0$ tension is greatly alleviated, and the $S_8$ tension is also slightly alleviated, which is difficult to realize in other cosmological models.

Although the $\tilde\Lambda$CDM model demonstrates potential in concurrently mitigating the $H_0$ and $S_8$ discrepancies, it leads to a larger AIC value compared to the $\Lambda$CDM model.
This indicates that compared to $\Lambda$CDM, the $\tilde\Lambda$CDM model's capability in fitting observational data is weaker for the CBS case.

\begin{figure}[t]
\includegraphics[trim=0cm 0.1cm 0cm 0cm,clip, width=0.45\textwidth]{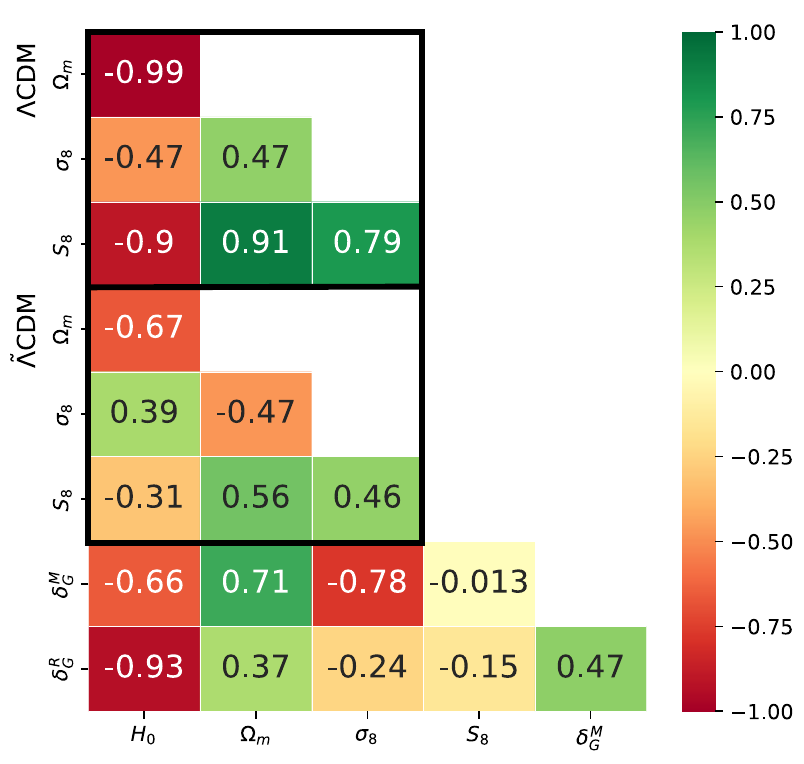}
\caption{Correlation coefficients between cosmological parameters in the $\Lambda$CDM and $\tilde\Lambda$CDM models using the CBS data.}
\label{cor}
\end{figure}

To ascertain the $\tilde\Lambda$CDM model's potential in reducing both the $H_0$ and $S_8$ tensions, we present the correlation coefficients between the cosmological parameters of the $\Lambda$CDM and $\tilde\Lambda$CDM models in the CBS case in Fig.~\ref{cor}.
Our findings reveal that within the $\tilde\Lambda$CDM model, the correlation coefficients of $\delta_G^M$ and $\delta_G^R$ with $H_0$ are negative, and the current constraints give negative $\delta_G^M$ and $\delta_G^R$ values.
Consequently, the $\tilde\Lambda$CDM model significantly alleviates the $H_0$ tension.
Additionally, the extra free parameters exhibit weak correlations with $S_8$, suggesting that mitigating the $S_8$ tension presents considerable challenges.

\begin{figure}
\centering
\begin{minipage}[t]{0.49\textwidth}
\centering
\includegraphics[width=\textwidth]{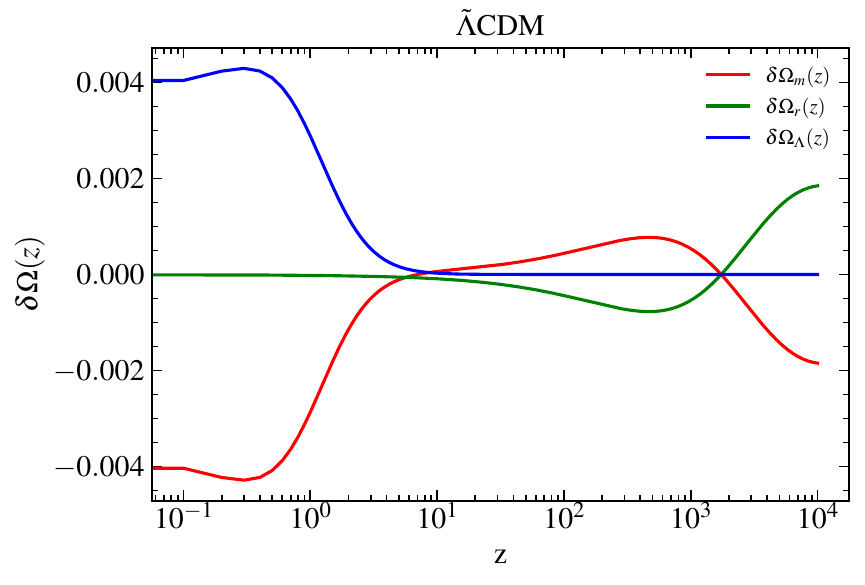}
\caption{
Differences $\delta\Omega_{{\rm m,r},\Lambda}(z)$ between the $\tilde\Lambda$CDM and $\Lambda$CDM models in the CBS case. Shown here is the best-fit result.}\label{Omega_tilde}
\end{minipage} \hfill
\end{figure}

The evolution of $\Omega_{{\rm m,r},\Lambda}$ with redshift $z$ in the $\tilde\Lambda$CDM model exhibits similarities to the $\Lambda$CDM counterparts.
To delineate the distinct characteristics between the $\tilde\Lambda$CDM and $\Lambda$CDM models, we define the differences as follows:
\begin{eqnarray}
\delta\Omega_{{\rm m,r},\Lambda}(z)=\Omega_{{\rm m,r},\Lambda}(z)-\Omega_{{\rm m,r},\Lambda}(z)|_{\Lambda{\rm CDM}}.
\label{diff}
\end{eqnarray}
{Employing the constrained results from the CBS dataset, as delineated in Table~\ref{tab1} and Fig.~\ref{CBSresults}, we graphically represent the differences in Fig.~\ref{Omega_tilde}.
As shown in Fig.~\ref{Omega_tilde}, the proportion of dark energy was nearly zero in the early universe, which is consistent with the result of the $\Lambda$CDM model. However, in the late universe, when dark energy begins to dominate cosmic evolution, its proportion is somewhat larger than that predicted by the $\Lambda$CDM model. Correspondingly, the proportion of matter density in the late universe is slightly less than that in the standard model. This characteristic means that in this model, the accelerated expansion of the late universe begins earlier and is more intense. Consequently, the age of the universe is younger than that predicted by the standard model, resulting in a larger Hubble constant.

As shown in Table~\ref{tab1} and Fig.~\ref{CBSresults}, the CBS data constrain the $\tilde\Lambda$CDM $H_0$ and $\sigma_8$ ($\Omega_m$), with the best-fit values larger (smaller) than their $\Lambda$CDM counterparts.
Nonetheless, it is evident that the confidence ranges of these parameters are more broad in the $\tilde\Lambda$CDM model compared to those in the $\Lambda$CDM model.
As a result, the $H_0$ tension relieves to $2.87\sigma$ in the $\tilde\Lambda$CDM model.
$S_8=\sigma_8(\Omega_{\rm m}/0.3)^{0.5}$ depends on both $\sigma_8$ and $\Omega_{\rm m}$. 
The $\tilde\Lambda$CDM $\Omega_m$ ($\sigma_8$) best-fit value decreases (increases) and its spreading increases, resulting in a decrease in
the $\tilde\Lambda$CDM $S_8$ tension to $2.77\sigma$, in contrast with the increasing $S_8$ tension found in many other models. 
Consequently, the interaction dynamics between dark energy and matter, as illustrated in Fig.~\ref{Omega_tilde}, are pivotal in substantially mitigating both the $H_0$ and $S_8$ tensions.}

\begin{figure*}[t]
\includegraphics[trim= 0.1cm 0.1cm 0cm 0cm,clip, width=0.6\textwidth]{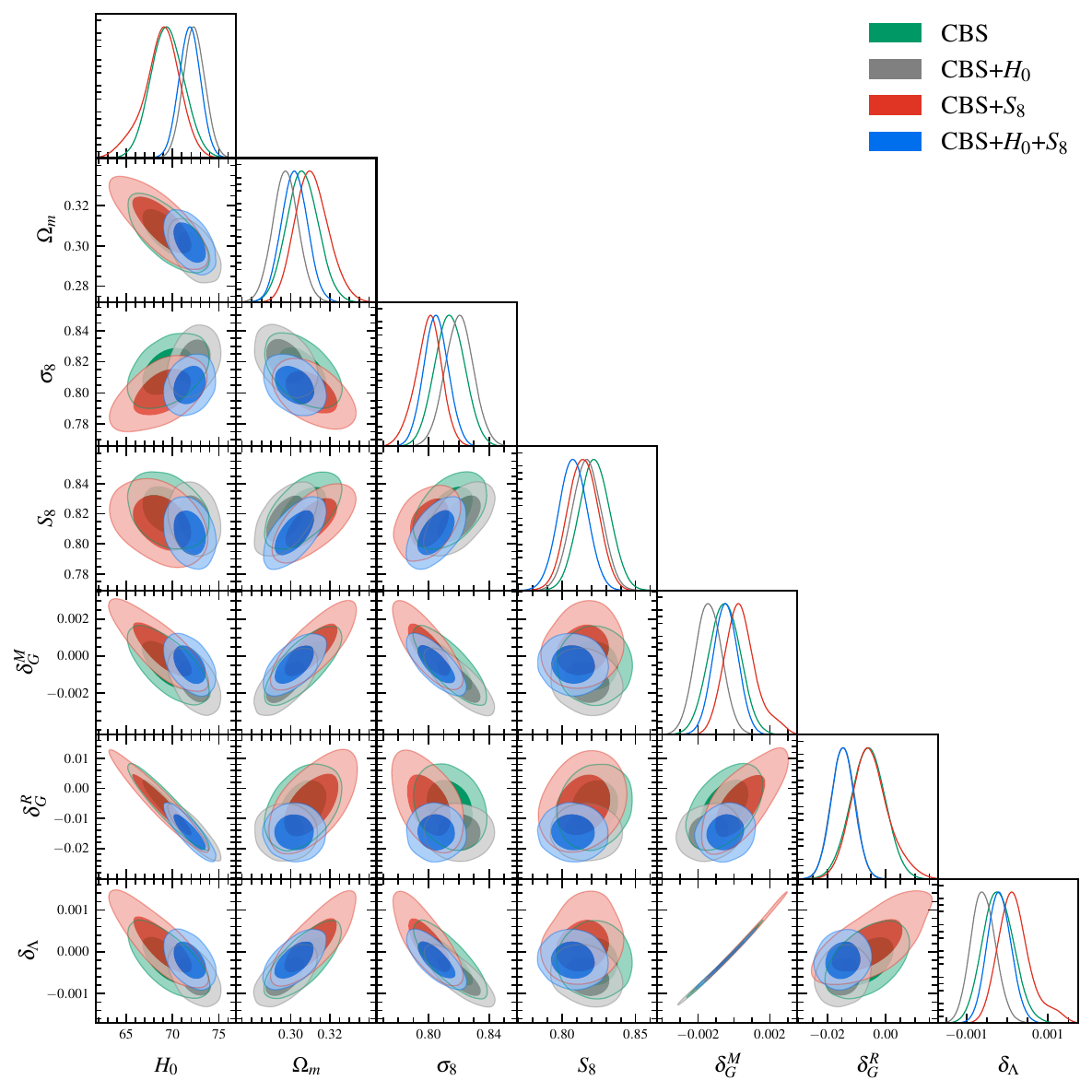}
\caption{Constraints ($68.3\%$ and $95.4\%$ confidence level) on the $\tilde\Lambda$CDM model using CBS, CBS+$H_0$, CBS+$S_8$, and CBS+$H_0$+$S_8$ data. $H_0$ is in units of ${\rm km~s^{-1}~Mpc^{-1}}$.} 
\label{alldatacon}
\end{figure*}

\begin{table*}
\renewcommand{\arraystretch}{1.3}%\tiny
\begin{tabular}{|c|c c c|}
\hline

Data &CBS+$H_0$&CBS+$S_8$&CBS+$H_0$+$S_8$\\

\hline
$\delta^M_{\rm G}$&$-0.00140\pm 0.00075$&$0.00038^{+0.00072}_{-0.00098}$&$-0.00046\pm 0.00069$\\

$\delta^R_{\rm G}$&$-0.0146\pm 0.0040$&$-0.0053^{+0.0055}_{-0.0068} $&$ -0.0147\pm 0.0040$\\

$\delta_{\Lambda}$&$-0.00059\pm 0.00030$&$0.00018^{+0.00031}_{-0.00046} $&$-0.00020\pm 0.00029 $\\

$\Omega_{\rm m}$&$0.2974\pm 0.0066$&$0.3108^{+0.0077}_{-0.0091} $&$0.3018\pm 0.0066 $\\

$H_0$&$ 72.3\pm 1.2$&$68.8^{+2.1}_{-1.7}$&$71.9\pm 1.2$\\

$\sigma_8$&$0.8203\pm 0.0095$&$ 0.8000^{+0.0093}_{-0.0079}$&$0.8050\pm 0.0081$\\
$S_8$&$0.817\pm 0.010$&$ 0.814\pm 0.010$&$ 0.8074\pm 0.0098$\\
\hline
${H_0}~{\rm tension}$&$0.47\sigma$&$1.96\sigma$&$0.72\sigma$\\
% ${\sigma_8}~{\rm tension}$&$2.43\sigma$&$1.84\sigma$&$2.11\sigma$\\
${S_8}~{\rm tension}$&$2.59\sigma$&$2.43\sigma$&$2.11\sigma$\\
$\chi_{\rm min}^{2}$&$1909.53$&$1912.46$&$1914.76$\\
% $\chi_{\rm min}^{2}$&$1909.46$&$1911.65$&$1914.74$\\
$\Delta {\rm AIC}$&$-2.07$&$5.02$&$-2.29$\\
\hline
\end{tabular}
\centering
\caption{Constraint results of parameters in the $\tilde\Lambda$CDM model with CBS+$H_0$, CBS+$S_8$, and CBS+$H_0$+$S_8$ data.}\label{tab2}
\end{table*}

Finally, we constrain the $\tilde\Lambda$CDM model using the three data combinations, CBS+$H_0$, CBS+$S_8$, and CBS+$H_0$+$S_8$. The constraint results are shown in Table~\ref{tab2} and Fig.~\ref{alldatacon}. 
We also compute the AIC values for the $\tilde\Lambda$CDM model using various data combinations to enable a systematic comparison between the $\tilde\Lambda$CDM and $\Lambda$CDM models across different datasets.
In the previous case (the CBS case), although the $H_0$ tension is greatly relieved and the $S_8$ tension is also slightly alleviated, the $\tilde\Lambda$CDM model is not favored by the CBS data because its $\Delta$AIC value is greater than 0 {($\Delta{\rm AIC}=3.79$)}. However, when the $H_0$ and $S_8$ data are added to the data combination, we find that the situation is significantly improved both in relieving the tensions and fitting the observations. 

In the CBS+$H_0$ case, we have {$H_0=72.3\pm 1.2$ ${\rm km~s^{-1}~Mpc^{-1}}$ and $S_8=0.817\pm 0.010$}.
Therefore, the $H_0$ and $S_8$ tensions are further relieved, and the $H_0$ tension is greatly relieved to {$0.47\sigma$}. Furthermore, in this case, {$\Delta{\rm AIC}=-2.07$}, indicating that the $\tilde\Lambda$CDM model is favored by the data over $\Lambda$CDM. However, the $\tilde\Lambda$CDM model is not favored by the CBS+$S_8$ data, in which though the $S_8$ tension is slightly improved {(to $2.43\sigma$)}, the {$H_0$ tension is slightly relieved (to $1.96\sigma$)}, and the fit gives {$\Delta{\rm AIC}=5.02$}. {Consequently, incorporating $H_0$ and $S_8$ priors into the data combination effectively mitigates the tensions.
Therefore, employing the CBS+$H_0$+$S_8$ dataset fosters an optimally synthesized scenario, for which we obtain $H_0=71.9\pm 1.2$ ${\rm km~s^{-1}~Mpc^{-1}}$ and $S_8=0.8074\pm 0.00998$, consequently reducing the $H_0$ and $S_8$ tensions to $0.72\sigma$ and $2.11\sigma$, respectively.
Furthermore, under these conditions, the model exhibits the highest degree of concordance with the data, as indicated by $\Delta{\rm AIC}=-2.29$.}

We compare our findings with those derived from the simplified $\tilde\Lambda$CDM model \cite{Gao:2021xnk}, wherein it is postulated that $\delta_{G}^M$ and $\delta_{G}^R$ are identical in Eqs.~(\ref{final+}) and (\ref{deltal+}), that is, $\delta_{G}^M=\delta_{G}^R=\delta_G$ and $\delta_\Lambda=\delta_G(\Omega_{\rm m}+\Omega_{\rm r})/\Omega_\Lambda$.
Employing the CBS dataset, the $H_0$ tension can be relieved to $3.59\sigma$ (with the best fit $H_0 = 67.71~{\rm km~s^{-1}~Mpc^{-1}}$), and in this case, the $S_8$ tension is exacerbated (with the best fit $S_8 = 0.8252$).
Additionally, within the simplified $\tilde\Lambda$CDM model constrained by the CBS + $H_0$ dataset, we obtain the best-fit values $\sigma_8=0.8720$ and $S_8=0.8310$, and the $\sigma_8$ and $S_8$ tensions significantly increase while alleviating the $H_0$ tension.

We extend our comparison to the $\Lambda(t)$CDM model, an interacting vacuum energy model that posits an energy exchange between vacuum energy and cold dark matter as detailed by the specific equations $\dot{\rho}_{\Lambda} = \beta H \rho_c$ and $\dot{\rho}_c+3H\rho_c = - \beta H \rho_c$ (here, the dimensionless parameter $\beta$ describes the interaction strength).
In Ref.~\cite{Guo:2018ans}, the same CBS+$H_0$ dataset is used to constrain the $\Lambda(t)$CDM model to obtain $H_0 = 69.36~{\rm km~s^{-1}~Mpc^{-1}}$ and $\sigma_8 = 0.844$.
In contrast, as illustrated in Table \ref{tab2}, the current findings explicitly show that the $\tilde\Lambda$CDM model exhibits distinct advantages over the $\Lambda(t)$CDM model.

%\noindent{\bf \em Conclusions.} 
{\it Conclusion.---}In this study, we propose a new cosmological model, in which the vacuum energy interacts with matter and radiation, which is considered to originate from the asymptotic safety and particle production of gravitational quantum field theory. We test this model using the current cosmological observations and discuss its capability for relieving the $H_0$ and $S_8$ tensions. 

To elucidate the mechanisms by which the ${\tilde\Lambda}$CDM model relieves tensions, we analyze the differences $\delta\Omega_{{\rm m,r},\Lambda}(z)$ between the ${\tilde\Lambda}$CDM and ${\Lambda}$CDM models.
The alleviation of the $H_0$ tension is attributed to the ${\tilde\Lambda}$CDM model's enhanced dark energy proportion and reduced matter fraction at low redshifts relative to the ${\Lambda}$CDM model, which stem from the conversion of matter into dark energy.
The findings of $\delta_G^{M, R} < 0$ and $\delta_\Lambda<0$ corroborate the $\tilde\Lambda$CDM hypothesis in which in the late universe radiation and matter decay into dark energy \cite{Xue2022, Xue:2023qft}.

We find that this cosmological model can significantly relieve the $H_0$ tension and simultaneously slightly reduce the $S_8$ tension, which cannot be easily observed in other cosmological models. When using the CBS data to constrain the model, we find that the $H_0$ tension is relieved to {$2.87\sigma$}, and meanwhile, the $S_8$ tension is also improved to {$2.77\sigma$}. However, in this case, the $S_8$ tension is only slightly reduced, and  the model is not favored by the CBS data {(because $\Delta{\rm AIC} = 3.79$)}.

When the $H_0$ and $S_8$ data are added to the data combination, the situation is significantly improved. In the CBS+$H_0$ case, we obtain the result {$H_0 = 72.3\pm 1.2$ ${\rm km~s^{-1}~Mpc^{-1}}$}, indicating that the $H_0$ tension is relieved to {$0.47\sigma$}, and in this case the model is favored over $\Lambda$CDM {($\Delta{\rm AIC} = -2.07$)}. In the CBS+$H_0$+$S_8$ case, we obtain a synthetically best situation, in which {$H_0 = 71.9\pm 1.2$ ${\rm km~s^{-1}~Mpc^{-1}}$ and $S_8 = 0.8074\pm 0.0098$}; thus, the $H_0$ and $S_8$ tensions are relieved to {$0.72\sigma$ and $2.11\sigma$}, respectively. In this case, the model is most favored by the data {($\Delta{\rm AIC} = -2.29$)}. 

Therefore, we find that such a cosmological model can greatly relieve the $H_0$ tension and simultaneously alleviate the $S_8$ tension.

Undoubtedly, this model requires further in-depth research in many aspects. Such an interaction between vacuum energy and matter is likely to introduce many additional observational effects. For example, we are not certain whether this model will lead to a significant integrated Sachs-Wolfe effect. Research on this issue requires further related theoretical study and analysis using full CMB angular power spectrum data. Moreover, this additional interaction may also lead to modifications in the middle-scale clustering patterns of large-scale structures, which will potentially affect the clustering strength at various scales. Therefore, such modifications may also affect the mass function and profile of dark matter halos, the statistical properties of galaxy clusters, the alignments of galaxies, the structure of the cosmic web, and so on. The study of these effects is a complicated issue, requires the use of N-body simulations, and also relies on assumptions about the nature of dark matter (such as cold or fuzzy dark matter). All these aspects deserve further in-depth discussion.

%\noindent{\bf \em Acknowledgements.} 
This work was supported by the National SKA Program of China (Grants Nos. 2022SKA0110200 and 2022SKA0110203) and the National Natural Science Foundation of China (Grants Nos. 11975072, 11875102, and 11835009).

%the science research grants from the China Manned Space Project (Grant No. CMS-CSST-2021-B01), the Liaoning Revitalization Talents Program (Grant No. XLYC1905011), the National Program for Support of Top-Notch Young Professionals (Grant No. W02070050), and the National 111 Project of China (Grant No. B16009).

%We are grateful for the support by the 111 Project (B16009), the National Natural Science Foundation of China (Grant Nos. 11975072, 11875102, 11835009, and 11690021), the Liaoning Revitalization Talents Program (Grant No. XLYC1905011), the Fundamental Research Funds for the Central Universities (Grant No. N2005030), and the Top-Notch Young Talents Program of China (Grant No. W02070050).

%%%REFERENCES%%%

\bibliographystyle{apsrev4-1.bst}
\bibliography{bbib}
 %the RSC's .bst file

% \onecolumngrid
% \appendix

% \section{APPENDIX}

%     \begin{table*}[h]
%     \renewcommand{\arraystretch}{1.3}%\tiny
%     \begin{tabular}{|c|c c c|}
%     \hline
    
%     Data &CBS+$H_0$&CBS+$S_8$&CBS+$H_0$+$S_8$\\
    
%     \hline
    
%     $\Omega_{\rm m}$&$ 0.3028\pm 0.0068$&$0.3086\pm 0.0055 $&$0.3021\pm 0.0052 $\\
    
%     $H_0$&$ 68.20\pm 0.60$&$67.74\pm 0.41$&$68.24\pm 0.40$\\
    
%     $\sigma_8$&$0.8093\pm 0.0065$&$0.8047\pm 0.0056 $&$ 0.8034\pm 0.0056$\\
%     $S_8$&$ 0.813\pm 0.013 $&$  0.816\pm 0.010$&$ 0.8062\pm 0.0098$\\
%     \hline
%     ${H_0}~{\rm tension}$&$4.03\sigma$&$4.74\sigma$&$4.31\sigma$\\
%     ${S_8}~{\rm tension}$&$2.20\sigma$&$2.54\sigma$&$2.05\sigma$\\
%     $\chi_{\rm min}^{2}$&$1915.60$&$1911.44$&$1921.05$\\

%     \hline
%     \end{tabular}
%     \centering
%     \caption{Constraint results of parameters in the $\Lambda$CDM model with CBS+$H_0$, CBS+$S_8$, and CBS+$H_0$+$S_8$ data.}\label{tab3}
%     \end{table*}

\end{document}